\documentclass[twocolumn,aps,pra]{revtex4-2}
\usepackage{bm,bbm}
\usepackage{amsfonts,amsmath,amssymb,latexsym}
\usepackage{graphicx}
\usepackage{mathtools}
\usepackage{xcolor}
\usepackage{hyperref}
\begin{document} 
\title{Accurate paraxial Maxwell beams}
\author{Tomasz Rado\.zycki}
\email{t.radozycki@uksw.edu.pl}
\affiliation{Faculty of Mathematics and Natural Sciences, College of Sciences, Institute of Physical Sciences, Cardinal Stefan Wyszy\'nski University, W\'oycickiego 1/3, 01-938 Warsaw, Poland} 

\begin{abstract}
In this work, the paraxial version of Maxwell equations is derived with the use of two Riemann-Silberstein vectors. Exact solutions of these equations are then obtained representing the paraxial electromagnetic fields. These fields satisfy the full Maxwell equations up to the order of $(\lambda/w_0)^2$. The solutions contain some additional terms, which turn out to be relevant for polarization, especially in the case of beams endowed with orbital angular momentum.
\end{abstract}

\maketitle

\section{Introduction}\label{intro}
Advances in sub-wavelength optics \cite{luo,neice}, for instance optical lithography \cite{ren,kaz}, optical quantum traps \cite{gaon,kolbow} or high-resolution microscopy (STED) \cite{sharma,sted}, etc. \!require a very precise characterization of the light beam near the optical axis. In this domain, the major technique for describing laser beams well collimated along the propagation axis constitutes the paraxial approximation~\cite{sie}. In such a circumstance the scalar approach is often acceptable, within which the electric field is postulated for example in the form of
\begin{equation}\label{epie}
\bm{E}(\bm{r},z,t)=\bm{E}_0 \Psi(\bm{r},z,t),
\end{equation}
with $\bm{E}_0$ standing for a constant vector. Then the envelope $\Psi(\bm{r},z,t)$ 
satisfies the so called paraxial equation~\cite{sie}:
\begin{equation}\label{paraxial}
\left(\mathcal{4}_\perp +2ik\partial_z\right) \psi({\bm{r}},z)=0,
\end{equation}
where $\mathcal{4}_\perp$ represents the two-dimensional Laplace operator acting in the transverse plane only. This assumption slightly violates for instance the Gauss law \cite{lax}, but is satisfactory for many purposes.

However, this approximate scalar description is inadequate when polarization effects of the wave become essential. Specifically, for highly focused beams, when the beam waist gets very small or only slightly larger than the wavelength, and the aperture half-angle grows, the conventional approach, which is either based on the scalar approximation or even incorporates the transverse components of the electric and magnetic fields, may be insufficient; the fully vectorial nature of the beam has to be accounted for, also including the longitudinal components of the fields. \cite{take,long}. 

Another important aspect are depolarization effects and polarization inhomogeneities in the perpendicular plane. They obviously cannot be neglected in super-precise applications like STED \cite{yang} and optical lithography \cite{bka}. Also of importance is the variation of polarization along the propagation axis, which affects the accurate manipulation of nanoparticles or atoms, especially for beams with complex structures such as Bessel-Gaussian or Laguerre-Gaussian beams. The modifications of polarization can then affect both gradient and scattering forces \cite{take,aloufi}.

Having this in mind, our aim in the present paper will be to obtain the most accurate and fully vectorial description of the laser beam within the framework of the paraxial approximation. It should be emphasized: the purpose of this work is not to obtain beyond-paraxial corrections to the fields derived under the ``arbitrary'' paraxial approximation -- as it is done elsewhere \cite{agra,dav,patta,shep,chen,yan,wang} -- but rather to first formulate a system of Maxwell paraxial equations and then solve them in a rigorous way. 

The paper is then organized as follows. In Sec. \ref{rsfields} the two Riemann-Silberstein (RS) \cite{web,sil} vectors are introduced in convenient variables. The approach that utilizes both complex RS vectors is preferable when the electromagnetic fields themselves are complex quantities. In Sec. \ref{parro} the paraxial Maxwell equations are established and in Sec. \ref{parmax} their accurate solutions are obtained. It is shown that, apart from the main terms usually accounted for, the exact paraxial expressions contain those of order $\varepsilon^2$ as well, where $\varepsilon\sim\lambda/w_0$  [for the definition of $\varepsilon$ see (\ref{deeps})]. In the subsequent Sec. \ref{acc} it is demonstrated through the investigation of the full (i.e., non-paraxial) Maxwell equations, that the latter are satisfied up to $\varepsilon^2$ terms, which is a better result than commonly accepted. Finally Sec. \ref{polari} is devoted to the influence of the additional terms on polarization and its inhomogeneities for beams with nonzero orbital angular momentum (OAM).

\section{Riemann-Silberstein vectors}\label{rsfields}

When deriving equations satisfied by the RS vectors, it is a handy choice to introduce complex variables in the plane perpendicular to the direction of wave propagation, as well as light-cone coordinates in place of $z$ and $t$:
\begin{subequations}\label{variab}
\begin{align}
& \sigma=x+iy,\qquad \bar{\sigma}=x-iy,\label{variaxy}\\
& \eta_-=z-ct,\qquad \eta_+=z+ct.\label{variaeta}
\end{align}
\end{subequations}
Denoting $\partial_\sigma=\partial/\partial\sigma$ and similarly for other variables, the derivatives with respect to $x,y,z,t$ can be expressed as
\begin{subequations}\label{dervar}
\begin{align}
& \partial_x=\partial_\sigma+\partial_{\bar{\sigma}},\qquad \partial_y=i(\partial_\sigma-\partial_{\bar{\sigma}}),\label{dervarxy}\\
& \partial_z=\partial_{\eta_+}+\partial_{\eta_-}\qquad \partial_t=c(\partial_{\eta_+}-\partial_{\eta_-}).\label{dervarzt}
\end{align}
\end{subequations}

As mentioned in Introduction in the case of complex electric and magnetic fields it is convenient to introduce two RS vectors defined as
\begin{subequations}\label{rsvectors}
\begin{align}
&\bm{F}=\bm{E}+ic\bm{B},\label{pluf}\\
&\bm{G}=\bm{E}-ic\bm{B},\label{plug}
\end{align}
\end{subequations}
and satisfying sourceless Maxwell equations:
\begin{subequations}\label{mfg}
\begin{align}
&\bm{\nabla}\times \bm{F}=\frac{i}{c}\,\partial_t\bm{F},\qquad \bm{\nabla}\bm{F}=0,\label{mff}\\
&\bm{\nabla}\times \bm{G}=-\frac{i}{c}\,\partial_t\bm{G},\qquad \bm{\nabla}\bm{G}=0,\label{mgg}
\end{align}
\end{subequations}
The electric and magnetic fields can be now directly recovered without necessity of taking the real and imaginary parts, which in the present case would be cumbersome:
\begin{subequations}\label{fieldsEB}
\begin{align}
&\bm{E}=\frac{1}{2}(\bm{F}+\bm{G}),\label{fieldsE}\\
&\bm{B}=-\frac{i}{2c}(\bm{F}-\bm{G}),\label{fieldsB}
\end{align}
\end{subequations}
The four equations (\ref{mff}) by means of (\ref{dervar}) can be given the form
\begin{subequations}\label{maxf}
\begin{align}
& i(\partial_\sigma-\partial_{\bar{\sigma}})F_z-(\partial_{\eta_+}+\partial_{\eta_-})F_y=i(\partial_{\eta_+}-\partial_{\eta_-})F_x,\label{maxfx}\\
& (\partial_{\eta_-}+\partial_{\eta_+})F_x-(\partial_\sigma+\partial_{\bar{\sigma}})F_z=i(\partial_{\eta_+}-\partial_{\eta_-})F_y,\label{maxfy}\\
& (\partial_\sigma+\partial_{\bar{\sigma}})F_y-i(\partial_\sigma-\partial_{\bar{\sigma}})F_x=i(\partial_{\eta_+}-\partial_{\eta_-})F_z,\label{maxfz}\\
& (\partial_\sigma+\partial_{\bar{\sigma}})F_x+i(\partial_\sigma-\partial_{\bar{\sigma}})F_y+(\partial_{\eta_+}+\partial_{\eta_-})F_z=0,\label{maxfxyz}
\end{align}
\end{subequations}
which can be simplified by combining (\ref{maxfx}) with (\ref{maxfy}) and  (\ref{maxfz})  with (\ref{maxfxyz}):
\begin{subequations}\label{mauf}
\begin{align}
& \partial_{\eta_+}(F_x-iF_y)=\partial_\sigma F_z,\label{maufx}\\
& \partial_{\eta_-}(F_x+iF_y)=\partial_{\bar{\sigma}} F_z,\label{maufy}\\
& \partial_\sigma(F_x+iF_y)=-\partial_{\eta_+} F_z,\label{maufz}\\
& \partial_{\bar{\sigma}}(F_x-iF_y)=-\partial_{\eta_-} F_z\label{maufxyz}
\end{align}
\end{subequations}
Acting with $\partial_{\bar{\sigma}}$ on (\ref{maufx}) and using (\ref{maufxyz}) [or alternatively acting with $\partial_\sigma$ on (\ref{maufy}) and using (\ref{maufz})] one obtains
\begin{equation}\label{eqwavf}
(\partial_{\bar{\sigma}}\partial_\sigma+\partial_{\eta_+}\partial_{\eta_-})F_z=0,
\end{equation}
which is nothing more than the wave equation. Naturally, the same equation is satisfied by the components $F_x$ and $F_y$, which can also be derived from set (\ref{mauf}) by simple manipulations.

Analogous operations can be made on the components of the Riemann-Silberstein vector $\bm G$. First we write down the set of Maxwell equations in the form similar to (\ref{maxf}):
\begin{subequations}\label{maxg}
\begin{align}
& i(\partial_\sigma-\partial_{\bar{\sigma}})G_z-(\partial_{\eta_+}+\partial_{\eta_-})G_y=-i(\partial_{\eta_+}-\partial_{\eta_-})G_x,\label{maxgx}\\
& (\partial_{\eta_-}+\partial_{\eta_+})G_x-(\partial_\sigma+\partial_{\bar{\sigma}})G_z=-i(\partial_{\eta_+}-\partial_{\eta_-})G_y,\label{maxgy}\\
& (\partial_\sigma+\partial_{\bar{\sigma}})G_y-i(\partial_\sigma-\partial_{\bar{\sigma}})G_x=-i(\partial_{\eta_+}-\partial_{\eta_-})G_z,\label{maxgz}\\
& (\partial_\sigma+\partial_{\bar{\sigma}})G_x+i(\partial_\sigma-\partial_{\bar{\sigma}})G_y+(\partial_{\eta_+}+\partial_{\eta_-})G_z=0\label{maxgxyz}
\end{align}
\end{subequations}
from which it can be obtained that
\begin{subequations}\label{maug}
\begin{align}
& \partial_{\eta_-}(G_x-iG_y)=\partial_\sigma G_z,\label{maugx}\\
& \partial_{\eta_+}(G_x+iG_y)=\partial_{\bar{\sigma}} G_z,\label{maugy}\\
& \partial_\sigma(G_x+iG_y)=-\partial_{\eta_-} G_z,\label{maugz}\\
& \partial_{\bar{\sigma}}(G_x-iG_y)=-\partial_{\eta_+} G_z,\label{maugxyz}
\end{align}
\end{subequations}
and finally
\begin{equation}\label{eqwavg}
(\partial_{\bar{\sigma}}\partial_\sigma+\partial_{\eta_+}\partial_{\eta_-})G_z=0.
\end{equation}
So far, the resultant sets (\ref{mauf}) and (\ref{maug}) do not involve any approximations. Instead, they are simply sets of Maxwell equations rewritten in useful variables, which provide a convenient starting point for obtaining the paraxial equations. This is handled in the following section.

\section{Paraxial equations}\label{parro}
 
In order to establish the paraxial version of the Maxwell equations for $\bm F$ and $\bm G$, we first detach the propagation factor $e^{ik(z-ct)}=e^{ik\eta_-}$, introducing the tilde vectors:
\begin{subequations}\label{mono}
\begin{align}
&\bm{F}(\sigma,\bar{\sigma},\eta_+,\eta_-)=e^{i k \eta_-}\widetilde{\bm{F}}(\sigma,\bar{\sigma},\eta_+)\label{monof}\\
&\bm{G}(\sigma,\bar{\sigma},\eta_+,\eta_-)=e^{i k \eta_-}\widetilde{\bm{G}}(\sigma,\bar{\sigma},\eta_+).\label{monog}
\end{align}
\end{subequations}
Now, as it is well known~\cite{hil93,trhan}, the paraxial approximation can be accomplished with the use of the following replacements in (\ref{mauf}) and (\ref{maug}) 
\begin{equation}\label{parp}
\eta_+\longmapsto 2z,\qquad \partial_{\eta_+}\longmapsto \frac{1}{2}\,\partial_z.
\end{equation}
The set of equations (\ref{mauf}) then becomes
\begin{subequations}\label{pauf}
\begin{align}
& \partial_z(\widetilde{F}_x-i\widetilde{F}_y)=2\partial_\sigma \widetilde{F}_z,\label{paufx}\\
& ik(\widetilde{F}_x+i\widetilde{F}_y)=\partial_{\bar{\sigma}} \widetilde{F}_z,\label{paufy}\\
& \partial_\sigma(\widetilde{F}_x+i\widetilde{F}_y)=-\frac{1}{2}\partial_z \widetilde{F}_z,\label{paufz}\\
& \partial_{\bar{\sigma}}(\widetilde{F}_x-i\widetilde{F}_y)=-ik \widetilde{F}_z.\label{paufxyz}
\end{align}
\end{subequations}
Simple manipulations on (\ref{paufx}) and  (\ref{paufxyz}) [or on (\ref{paufy}) and  (\ref{paufz})] lead to the conclusion that each component of $\widetilde{\bm{F}}$ actually satisfies the scalar paraxial equation (for brevity written below only for $\widetilde{F}_z$)
\begin{equation}\label{pareqf}
\left(\partial_{\bar{\sigma}}\partial_\sigma+\frac{1}{2}\,ik\partial_z\right)\widetilde{F}_z=0,\;\;\;\; \mathrm{i.e.} \;\;\;\; \left(\mathcal{4}_\perp+2ik\partial_z\right)\widetilde{F}_z=0.
\end{equation}
This explicitly demonstrates the correctness of the the substitutions (\ref{parp}). 
Following the same steps with the system (\ref{maug}), one gets
\begin{subequations}\label{paug}
\begin{align}
& \partial_z(\widetilde{G}_x+i\widetilde{G}_y)=2\partial_\sigma \widetilde{G}_z,\label{paugx}\\
& ik(\widetilde{G}_x-i\widetilde{G}_y)=\partial_{\bar{\sigma}} \widetilde{G}_z,\label{paugy}\\
& \partial_{\bar{\sigma}}(\widetilde{G}_x-i\widetilde{G}_y)=-\frac{1}{2}\partial_z \widetilde{G}_z,\label{paugz}\\
& \partial_\sigma(\widetilde{G}_x+i\widetilde{G}_y)=-ik \widetilde{G}_z,\label{paugxyz}
\end{align}
\end{subequations}
and consequently
\begin{equation}\label{pareqg}
\left(\partial_\sigma\partial_{\bar{\sigma}}+\frac{1}{2}\,ik\partial_z\right)\widetilde{G}_z=0,\;\;\;\; \mathrm{i.e.} \;\;\;\;  \left(\mathcal{4}_\perp+2ik\partial_z\right)\widetilde{G}_z=0.
\end{equation}
As the RS vectors are linear combinations of $\bm{E}$ and $\bm{B}$, it is obvious that the paraxial equation will be equally obeyed by all the components of the tilde electric and magnetic fields obtained from $\widetilde{\bm{F}}$ and $\widetilde{\bm{G}}$ through the relations parallel to (\ref{fieldsEB}).

In order to formulate the paraxial Maxwell equations in terms of $\widetilde{\bm{E}}$ and $\widetilde{\bm{B}}$ in the compact form, let us introduce the paraxial nabla operator
\begin{equation}\label{pargra}
\widetilde{\bm{\nabla}}=\left[\partial_x,\partial_y,ik+\frac{1}{2}\,\partial_z\right],
\end{equation}
while the regular operator (taking into account (\ref{mono})) has the form
\begin{equation}\label{gra}
\bm{\nabla}=\left[\partial_x,\partial_y,ik+\partial_z\right].
\end{equation}
Manipulating equations (\ref{pauf}) and (\ref{paug}) one gets in the straightforward way the following set of sourceless paraxial Maxwell equations for electromagnetic fields:
\begin{subequations}\label{maxwellEB}
\begin{align}
\widetilde{\bm{\nabla}}\times\widetilde{\bm{E}}&=c\left(ik-\frac{1}{2}\,\partial_z\right)\widetilde{\bm{B}},\label{maxwellE}\\
\widetilde{\bm{\nabla}}\times\widetilde{\bm{B}}&=-\frac{1}{c}\left(ik-\frac{1}{2}\,\partial_z\right)\widetilde{\bm{E}},\label{maxwellB}\\
\widetilde{\bm{\nabla}}\widetilde{\bm{E}}&=0,\label{maxwellGE}\\
\widetilde{\bm{\nabla}}\widetilde{\bm{B}}&=0.\label{maxwellGB}
\end{align}
\end{subequations}
The set (\ref{maxwellEB}) can provide the starting point for determining paraxial monochromatic fields with a distinguished propagation axis $z$. This is worked out in the subsequent section, although instead of solving (\ref{maxwellEB}), we prefer to deal with (\ref{pauf}) and (\ref{paug}), which constitutes an alternative way of proceeding.

\section{Solutions to the paraxial Maxwell equations}\label{parmax}

In order to find the solutions of (\ref{pauf}), (\ref{paug}), or equivalently (\ref{maxwellEB}), we first introduce two scalars: $\widetilde{\Phi}$ and $\widetilde{\Psi}$, that are primitive functions for $\widetilde{F}_z$ and $\widetilde{G}_z$ with respect to the variable $z$, i.e.,
\begin{subequations}\label{phipsi}
\begin{align}
&\widetilde{\Phi}(\sigma,\bar{\sigma},z)=\int\! dz\, \widetilde{F}_z(\sigma,\bar{\sigma},z),\label{phidef}\\
&\widetilde{\Psi}(\sigma,\bar{\sigma},z)=\int\! dz\, \widetilde{G}_z(\sigma,\bar{\sigma},z),\label{psidef}
\end{align}
\end{subequations}
Then equations (\ref{paufx}) and (\ref{paufy}) can be solved for other components of $\widetilde{\bm{F}}$, giving
\begin{subequations}\label{rof}
\begin{align}
&\widetilde{F}_x=\left(\partial_\sigma-\frac{i}{2k}\,\partial_z\partial_{\bar{\sigma}}\right)\widetilde{\Phi},\label{rofx}\\
&\widetilde{F}_y=i\left(\partial_\sigma+\frac{i}{2k}\,\partial_z\partial_{\bar{\sigma}}\right)\widetilde{\Phi},\label{rofy}\\
&\widetilde{F}_z=\partial_z\widetilde{\Phi}.\label{rofz}
\end{align}
\end{subequations}

Mutatis mutandis all the components of $\widetilde{\bm{G}}$ are expressed through $\widetilde{\Psi}$:
\begin{subequations}\label{rog}
\begin{align}
&\widetilde{G}_x=\left(\partial_{\bar{\sigma}}-\frac{i}{2k}\,\partial_z\partial_\sigma\right)\widetilde{\Psi},\label{rogx}\\
&\widetilde{G}_y=-i\left(\partial_{\bar{\sigma}}+\frac{i}{2k}\,\partial_z\partial_\sigma\right)\widetilde{\Psi},\label{rogy}\\
&\widetilde{G}_z=\partial_z\widetilde{\Psi}.\label{rogz}
\end{align}
\end{subequations}
Both $\widetilde{\Phi}$ and $\widetilde{\Psi}$ have to be solutions of the scalar paraxial equation (\ref{pareqf}) or (\ref{pareqg}), which ensures the fulfillment of equations (\ref{paufz}) and (\ref{paufxyz}), as well as  (\ref{paugz}) and (\ref{paugxyz}) respectively. Indeed, up to a constant which can be set to zero we have:
\begin{eqnarray}
\partial_{\bar{\sigma}}\partial_\sigma\widetilde{\Phi}&=&\int\! dz\, \partial_{\bar{\sigma}}\partial_\sigma\widetilde{F}_z(\sigma,\bar{\sigma},z)\label{scalap}\\
&=&-\frac{1}{2}\,ik\int\! dz\, \partial_z\widetilde{F}_z(\sigma,\bar{\sigma},z)=-\frac{1}{2}\,ik\partial_z \widetilde{\Phi},\nonumber
\end{eqnarray}
 and the same for $\widetilde{\Psi}$:
\begin{eqnarray}
\partial_\sigma\partial_{\bar{\sigma}}\widetilde{\Psi}&=&\int\! dz\, \partial_\sigma\partial_{\bar{\sigma}}\widetilde{G}_z(\sigma,\bar{\sigma},z)\label{scalapa}\\
&=&-\frac{1}{2}\,ik\int\! dz\, \partial_z\widetilde{G}_z(\sigma,\bar{\sigma},z)=-\frac{1}{2}\,ik\partial_z \widetilde{\Psi}.\nonumber
\end{eqnarray}

These two scalar functions may be treated as certain ``potentials'' playing the role similar as Whittaker functions do \cite{whittaker}, but it is more appropriate to define as the fundamental potentials  the following combinations:
\begin{equation}\label{vpm}
\widetilde{V}_\pm=\frac{1}{4}\left(\widetilde{\Phi}\pm\widetilde{\Psi}\right).
\end{equation}
This allows to write down in a fairly simple manner the expressions of the electric field components:
\begin{subequations}\label{parpe}
\begin{align}
&\widetilde{E}_x=\left(1-\frac{i}{2k}\,\partial_z\right)\partial_x\widetilde{V}_+-i\left(1+\frac{i}{2k}\,\partial_z\right)\partial_y\widetilde{V}_-,\label{parpex}\\
&\widetilde{E}_y=i\left(1+\frac{i}{2k}\,\partial_z\right)\partial_x\widetilde{V}_-+\left(1-\frac{i}{2k}\,\partial_z\right)\partial_y\widetilde{V}_+,\label{parpey}\\
&\widetilde{E}_z=2\partial_z\widetilde{V}_+.\label{parpez}
\end{align}
\end{subequations}
and those of the magnetic field:
\begin{subequations}\label{parpb}
\begin{align}
&\widetilde{B}_x=-\frac{1}{c}\left[i\left(1-\frac{i}{2k}\,\partial_z\right)\partial_x\widetilde{V}_-+\left(1+\frac{i}{2k}\,\partial_z\right)\partial_y\widetilde{V}_+\right],\label{parpbx}\\
&\widetilde{B}_y=\frac{1}{c}\left[\left(1+\frac{i}{2k}\,\partial_z\right)\partial_x\widetilde{V}_+-i\left(1-\frac{i}{2k}\,\partial_z\right)\partial_y\widetilde{V}_-\right],\label{parpby}\\
&\widetilde{B}_z=-\frac{2i}{c}\,\partial_z\widetilde{V}_-,\label{parpbz}
\end{align}
\end{subequations}
Equations (\ref{scalap}), (\ref{scalapa}) together with (\ref{vpm}) ensure that each of the components of the fields $\widetilde{\bm{E}}$ and $\widetilde{\bm{B}}$ obeys the scalar paraxial equation. Naturally, one can also ascertain by a bit longer but straightforward calculation that these fields satisfy Maxwell paraxial equations (\ref{maxwellEB}).

\section{Accuracy of the paraxial expressions}\label{acc}

Since the electromagnetic fields in the form given by (\ref{parpe}) and (\ref{parpb}) are found within the paraxial approximation, obviously they do not meet the exact Maxwell equations. Therefore, it is essential to establish the terms that violate these equations and to estimate their magnitude in a realistic situation of a laser light beam. Such a feasible beam  is characterized, among others, by two parameters: the radius of the beam waist denoted with $w_0$ and the so-called Rayleigh length, i.e. the distance measured along the axis of propagation, at which the cross-section of the beam doubles: $z_R=kw_0^2/2$. These quantities establish the scales of distance at which the wave fields in the transverse ($w_0$) and longitudinal ($z_R$) directions change in a significant way. For this reason, it is convenient, for the purposes of further discussion, to make use of the dimensionless coordinates defined as
\begin{equation}\label{bezw}
\xi_x=\frac{x}{w_0},\qquad \xi_y=\frac{y}{w_0},\qquad \zeta=\frac{z}{z_R}.
\end{equation}
In a typical situation the following inequalities hold
\begin{equation}\label{defy}
\lambda\ll w_0\ll z_R=\frac{kw_0^2}{2},
\end{equation}
which lets us define a dimensionless small parameter
\begin{equation}\label{deeps}
\varepsilon=\frac{1}{kw_0}=\frac{\lambda}{2\pi w_0}=\frac{w_0}{2 z_R},
\end{equation}
which represents also the aperture half-angle: $\theta=2\varepsilon$.
The value of this parameter is usually indeed tiny and, for example, for a red beam of wavelength $\lambda \approx 650\,\mathrm{nm}$ and $w_0=1\,\mathrm{mm}$, one finds $\varepsilon\approx 10^{-4}$. However, in ultra-high-precision optics $w_0$ can be reduced to very small values (in extreme cases even to sub-wavelength ones), which would make $\varepsilon$ to become as large as $0.2$ (see for instance \cite{young,linma}).  

It is also very handy to absorb the constants $w_0$ and light velocity $c$ into the electromagnetic fields as follows (alternatively one might redefine functions $\widetilde{V}_\pm$)
\begin{equation}\label{bezeb}
\widetilde{\bm{\mathcal{E}}}=w_0 \widetilde{\bm{E}},\qquad \widetilde{\bm{\mathcal{B}}}=cw_0 \widetilde{\bm{B}}.
\end{equation}
For the sake of the following estimates, these fields can be considered dimensionless, which corresponds to the choice of the dimensionless potentials $\widetilde{V}_\pm$. Any dimensionful constants can be easily reinstated in (\ref{parpe}) and (\ref{parpb}).

Let us now rewrite the fields (\ref{parpe}) and  (\ref{parpb}) in this dimensionless form, inserting an additional parameter $\beta$, which ultimately will be set to $1$. The presence of this coefficient helps us to establish the role of the $\varepsilon^2$ terms that are often omitted in paraxial fields~(for instance \cite{agra,dav,patta,chen,mukunda,sim}). Although our derivation in the previous sections ensures that $\beta=1$, for any other value the components of the electric and magnetic fields will also fulfill the paraxial equation, since $\widetilde{V}_\pm$ potentials do [but, of course, (\ref{maxwellEB}) cease to be satisfied]. Below, when examining full Maxwell equations below, it will become clear that the value found, i.e., $\beta=1$ is optimal.

The components of the dimensionless electric field then take temporarily the form
\begin{subequations}\label{parpee}
\begin{align}
\;&\widetilde{\mathcal{E}}_{\xi_x}=\partial_{\xi_x}\left(1-i\beta\varepsilon^2\partial_\zeta \right)\widetilde{V}_+-i\partial_{\xi_y}\left(1+i\beta\varepsilon^2\partial_\zeta\right)\widetilde{V}_-,\label{parpexe}\\
&\widetilde{\mathcal{E}}_{\xi_y}=i\partial_{\xi_x}\left(1+i\beta\varepsilon^2\partial_\zeta\right)\widetilde{V}_-+\partial_{\xi_y}\left(1-i\beta\varepsilon^2\partial_\zeta\right)\widetilde{V}_+,\label{parpeye}\\
&\widetilde{\mathcal{E}}_{\zeta}=4\varepsilon\partial_\zeta\widetilde{V}_+,\label{parpeze}
\end{align}
\end{subequations}
and similarly for the magnetic field
\begin{subequations}\label{parpbe}
\begin{align}
&\widetilde{\mathcal{B}}_{\xi_x}=-i\partial_{\xi_x}\left(1-i\beta\varepsilon^2\partial_\zeta\right)\widetilde{V}_--\partial_{\xi_y}\left(1+i\beta\varepsilon^2\partial_\zeta\right)\widetilde{V}_+,\label{parpbxe}\\
&\widetilde{\mathcal{B}}_{\xi_y}=\partial_{\xi_x}\left(1+i\beta\varepsilon^2\partial_\zeta\right)\widetilde{V}_+-i\partial_{\xi_y}\left(1-i\beta\varepsilon^2\partial_\zeta\right)\widetilde{V}_-,\label{parpbye}\\
&\widetilde{\mathcal{B}}_{\zeta}=-4i\varepsilon\partial_\zeta\widetilde{V}_-.\label{parpbze}
\end{align}
\end{subequations}

Now we are going to find the perturbations that these fields entail in the exact Maxwell equations. All these equations will be given the form that the breaking terms will be placed on the r.h.s, i.e., those that would vanish if the equations were satisfied exactly. To this goal let us introduce the dimensionless nabla operator (\ref{gra}):
\begin{equation}\label{nabbe}
\widehat{\bm{\nabla}}=\left[\partial_{\xi_x},\partial_{\xi_y},\frac{i}{\varepsilon}+2\varepsilon\partial_\zeta\right].
\end{equation}
The appearance of the term  $i/\varepsilon$ signals that the extra terms in (\ref{parpee}) and (\ref{parpbe}), i.e., those proportional to $\varepsilon^2$, will actually prove necessary and cannot be omitted or modified. 

Upon substitution of (\ref{parpee}) and (\ref{parpbe}) into the two ``divergence equations'' one gets:
\begin{subequations}\label{ndok}
\begin{align}
&\widehat{\bm{\nabla}}\widetilde{\bm{\mathcal{E}}}=4(2-\beta)\varepsilon^2\partial^2_\zeta\widetilde{V}_+,\label{nkoker}\\
&\widehat{\bm{\nabla}}\widetilde{\bm{\mathcal{B}}}=-4i(2-\beta)\varepsilon^2\partial^2_\zeta\widetilde{V}_-,\label{nkokb}
\end{align}
\end{subequations}
When setting $\beta=2$ it is possible to make these two equations be satisfied by our paraxial fields in an exact manner.  However, the accuracy of the other equations, that is, the ``rotation equations'' (Faraday's and Amp\`ere's), would suffer, as can be seen from the following calculations: 
\begin{eqnarray}
\widehat{\bm{\nabla}}&\times&\widetilde{\bm{\mathcal{E}}}-\frac{i}{\varepsilon}\widetilde{\bm{\mathcal{B}}}\label{piea}\\
&=&2(1-\beta)\varepsilon\partial_\zeta\left[\partial_{\xi_y}\widetilde{V}_+-i\partial_{\xi_x}\widetilde{V}_-,-\partial_{\xi_x}\widetilde{V}_+-i\partial_{\xi_y}\widetilde{V}_-,0\right]\nonumber\\
&+&4i\beta\varepsilon^2\partial_\zeta^2\left[0,0,\widetilde{V}_-\right]\nonumber
\\
&+&2\beta\varepsilon^3\partial_\zeta^2\left[\partial_{\xi_x}\widetilde{V}_-+i\partial_{\xi_y}\widetilde{V}_+,-i\partial_{\xi_x}\widetilde{V}_++\partial_{\xi_y}\widetilde{V}_-,0\right],\nonumber
\end{eqnarray}
and
\begin{eqnarray}
\widehat{\bm{\nabla}}&\times&\widetilde{\bm{\mathcal{B}}}+\frac{i}{\varepsilon}\widetilde{\bm{\mathcal{E}}}\label{piba}\\
&=&2(1-\beta)\varepsilon\partial_\zeta\left[i\partial_{\xi_y}\widetilde{V}_--\partial_{\xi_x}\widetilde{V}_+,i\partial_{\xi_x}\widetilde{V}_--\partial_{\xi_y}\widetilde{V}_+,0\right]\nonumber\\
&+&4\beta\varepsilon^2\partial_\zeta^2\left[0,0,\widetilde{V}_+\right]\nonumber
\\
&+&2\beta\varepsilon^3\partial_\zeta^2\left[-i\partial_{\xi_x}\widetilde{V}_++\partial_{\xi_y}\widetilde{V}_-,-\partial_{\xi_x}\widetilde{V}_--i\partial_{\xi_y}\widetilde{V}_+,0\right].\nonumber
\end{eqnarray}
The expressions on the right-hand sides of (\ref{piea}) and (\ref{piba}) have been collected following the powers of the parameter $\varepsilon$. It is obvious that in order to ascertain the fulfillment of all Maxwell equations with accuracy up to $\varepsilon^2$ one cannot have $\beta=0$, as is often assumed, but rather $\beta=1$, which also follows from our former derivation. The extra terms in (\ref{parpee}) and (\ref{parpbe}), i.e. in (\ref{parpe}) and (\ref{parpb}) prove then to be necessary in the obtained form.

One remark should be made here: these supplementary terms are not beyond-paraxial corrections to paraxial solutions, such as for instance those in~\cite{agra,dav,patta,shep,chen,yan,wang}. On the contrary, the expressions for $\widetilde{\bm{\mathcal{E}}}$ and $\widetilde{\bm{\mathcal{B}}}$ represent ``accurate'' paraxial fields, which eventually can be subject to further non-paraxial perturbation calculations. This was stressed in Introduction

In what follows the value $\beta=1$ is reinstated and one finally obtains all Maxwell equations satisfied up to $\varepsilon^2$ terms:
\begin{subequations}\label{romax}
\begin{align}
&\widehat{\bm{\nabla}}\widetilde{\bm{\mathcal{E}}}=4\varepsilon^2\partial^2_\zeta\widetilde{V}_+,\label{romax1}\\
&\widehat{\bm{\nabla}}\widetilde{\bm{\mathcal{B}}}=-4i\varepsilon^2\partial^2_\zeta\widetilde{V}_-,\label{romax2}\\
&\widehat{\bm{\nabla}}\times\widetilde{\bm{\mathcal{E}}}-\frac{i}{\varepsilon}\widetilde{\bm{\mathcal{B}}}=2\varepsilon^2\partial_\zeta^2\Big[\varepsilon\left(\partial_{\xi_x}\widetilde{V}_-+i\partial_{\xi_y}\widetilde{V}_+\right),\label{romax3}\\
&\hspace{23ex}\varepsilon\left(-i\partial_{\xi_x}\widetilde{V}_++\partial_{\xi_y}\widetilde{V}_-\right),2i\widetilde{V}_-\Big]\nonumber\\
&\widehat{\bm{\nabla}}\times\widetilde{\bm{\mathcal{B}}}+\frac{i}{\varepsilon}\widetilde{\bm{\mathcal{E}}}=
2\varepsilon^2\partial_\zeta^2\Big[\varepsilon\left(-\partial_{\xi_x}\widetilde{V}_++\partial_{\xi_y}\widetilde{V}_-\right),\label{romax4}\\
&\hspace{23ex}\varepsilon\left(-\partial_{\xi_x}\widetilde{V}_--i\partial_{\xi_y}\widetilde{V}_+\right),2\widetilde{V}_+\Big].\nonumber
\end{align}
\end{subequations}

\section{Polarization effects}\label{polari}
\subsection{Azimuthal polarization}\label{azpol}

Below we are going to deal with cylindrical beams and make use of dimensionless fields (\ref{parpee}) and (\ref{parpbe}), naturally with $\beta$ reset to unity as stated above. Rewriting fields in cylindrical coordinates $\xi,\phi, \zeta$, one gets
\begin{subequations}\label{parpexfa}
\begin{align}
&\widetilde{\mathcal{E}}_\xi=\partial_\xi\left(1-i\varepsilon^2\partial_\zeta \right)\widetilde{V}_+-\frac{i}{\xi}\,\partial_{\phi}\left(1+i\varepsilon^2\partial_\zeta\right)\widetilde{V}_-,\label{parpexi}\\
&\widetilde{\mathcal{E}}_\phi=i\partial_\xi\left(1+i\varepsilon^2\partial_\zeta\right)\widetilde{V}_-+\frac{1}{\xi}\,\partial_{\phi}\left(1-i\varepsilon^2\partial_\zeta\right)\widetilde{V}_+,\label{parpefi}\
\end{align}
\end{subequations}
and
\begin{subequations}\label{parpbxf}
\begin{align}
&\widetilde{\mathcal{B}}_\xi=-i\partial_\xi\left(1-i\varepsilon^2\partial_\zeta \right)\widetilde{V}_--\frac{1}{\xi}\,\partial_{\phi}\left(1+i\varepsilon^2\partial_\zeta\right)\widetilde{V}_+,\label{parpbxi}\\
&\widetilde{\mathcal{B}}_\phi=\partial_\xi\left(1+i\varepsilon^2\partial_\zeta\right)\widetilde{V}_+-\frac{i}{\xi}\,\partial_{\phi}\left(1-i\varepsilon^2\partial_\zeta\right)\widetilde{V}_-,\label{parpbfi}
\end{align}
\end{subequations}
with $\zeta$ components given by (\ref{parpeze}) and (\ref{parpbze}). It is relatively simple to produce a beam with azimuthal polarization which is constant along $\zeta$. One can choose $\widetilde{V}_+=0$, and $\widetilde{V}_-$ to be $\phi$ independent, as it occurs with a fundamental Gaussian beam, i.e.,
\begin{equation}\label{fgb}
\widetilde{V}_-(\xi, \phi,\zeta)=\frac{1}{1+i\zeta}\, e^{-\frac{\xi^2}{1+i\zeta}},
\end{equation} 
where a dimensionful constant coefficient is omitted (and likewise below). Then one finds
\begin{subequations}\label{parpexf}
\begin{align}
&\widetilde{\mathcal{E}}_\xi=0,\quad \widetilde{\mathcal{E}}_\phi=i\partial_\xi\left(1+i\varepsilon^2\partial_\zeta\right)\widetilde{V}_-,\quad\widetilde{\mathcal{E}}_\zeta=0,\label{azyze}\\
&\widetilde{\mathcal{B}}_\xi=-i\partial_\xi\left(1-i\varepsilon^2\partial_\zeta \right)\widetilde{V}_-,\quad \widetilde{\mathcal{B}}_\phi=0,\quad \widetilde{\mathcal{B}}_\zeta=-4i\varepsilon\partial_\zeta\widetilde{V}_-. \label{bzyze}
\end{align}
\end{subequations}

\begin{figure}[h!]
\begin{center}
\includegraphics[width=0.48\textwidth,angle=0]{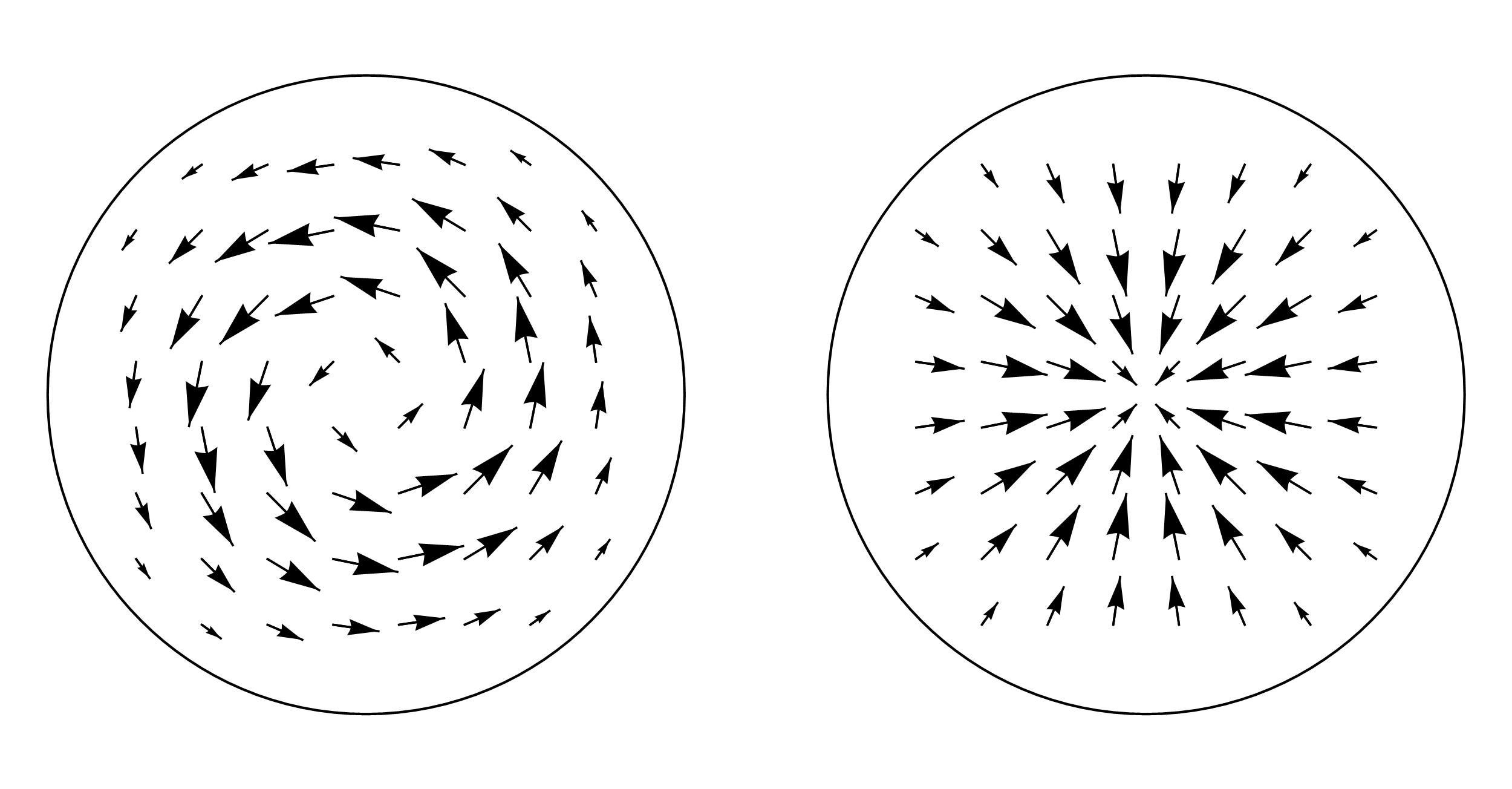}
\end{center}
\vspace{-4.5ex}
\caption{Vectors of electric field (\ref{azyze}) with the reinstated factor of $\exp\left({\frac{i}{2\varepsilon^2}\,\zeta}\right)$  [see (\ref{mono})] for $\varepsilon=0.2$ and $\zeta=0.5$ -- left diagram, and the same for magnetic field (\ref{bzyze}) -- right diagram, viewed from the wave front. The size of the arrows reflects the value of the electromagnetic fields at a given point.  The size of the circles corresponds to the maximum value of $\xi=1.4$, which amounts to $1.4 w_0$.}
\label{EBazy1}
\end{figure}

The electric field is purely transverse, and the longitudinal component of $\widetilde{\bm{\mathcal{B}}}$ is of order $\varepsilon$, i.e., increases for tightly focused beam. The vectors representing the real parts of $\widetilde{\bm{\mathcal{E}}}$ and $\widetilde{\bm{\mathcal{B}}}$ are depicted in Fig.~\ref{EBazy1} (the latter has the small longitudinal part as well, which is not visualized).

Surely one can choose the opposite: to set $\widetilde{V}_-=0$ and $\widetilde{V}_+$ in the form of (\ref{fgb}). Then the fields swap their roles:
\begin{subequations}\label{parpexb}
\begin{align}
&\widetilde{\mathcal{E}}_\xi=\partial_\xi\left(1-i\varepsilon^2\partial_\zeta \right)\widetilde{V}_+,\quad \widetilde{\mathcal{E}}_\phi=0,\quad\widetilde{\mathcal{E}}_\zeta=4\varepsilon\partial_\zeta\widetilde{V}_+,\label{azyzb}\\
&\widetilde{\mathcal{B}}_\xi=0,\quad \widetilde{\mathcal{B}}_\phi=\partial_\xi\left(1+i\varepsilon^2\partial_\zeta\right)\widetilde{V}_+,\quad \widetilde{\mathcal{B}}_\zeta=0, \label{bzyzb}
\end{align}
\end{subequations}
It is impossible, however, to generate the purely radial polarization that way, since Gaussian factors cannot be assumed to be $\zeta$ independent.

\subsection{Quasilinear polarization}\label{qlinpol}

\begin{figure}[h!]
\begin{center}
\includegraphics[width=0.48\textwidth,angle=0]{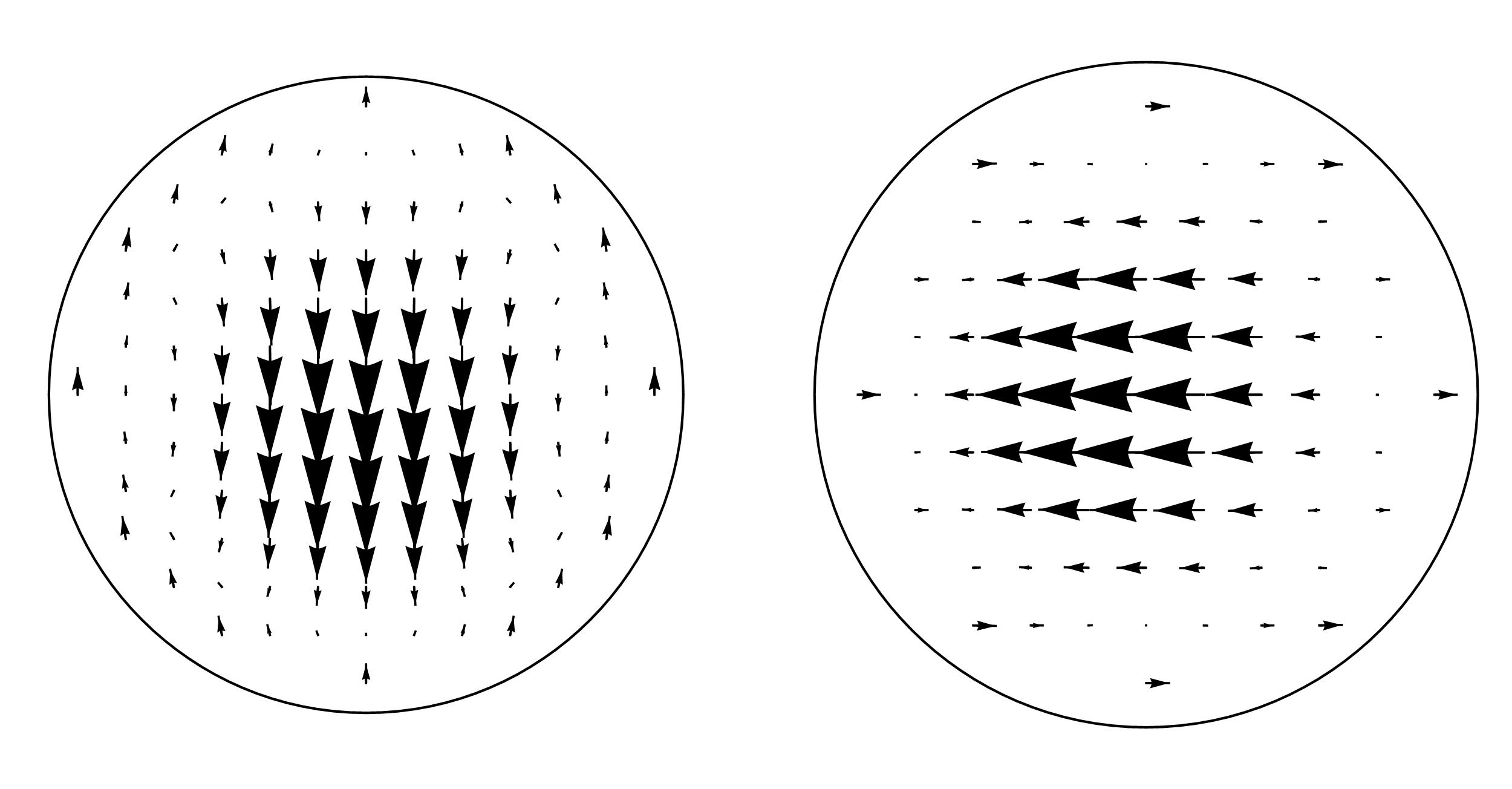}
\end{center}
\vspace{-4.5ex}
\caption{Vectors of electric field (\ref{parline}) with the reinstated factor of $\exp\left({\frac{i}{2\varepsilon^2}\,\zeta}\right)$  for $\varepsilon=0.2$ and $\zeta=0.8$ -- left diagram, and the same for polarization obtained from (\ref{lipara}) -- right diagram, viewed from the wave front. The size of the arrows reflects the value of the electric field at a given point. The appearance of the specific field distribution depends on the cross-section selected (i.e. the value of $\zeta$), but the general characteristics are the same. The size of the circles corresponds to the maximum value of $\xi=1.1$, which amounts to $1.1 w_0$.}
\label{Elin}
\end{figure}

To look for a beam with linear polarization, let us choose the potentials $\widetilde{V}_\pm$ in the form of superpositions of two scalar Gaussian beams of order $\pm 1$ (i.e., with OAM values equal to $\pm 1$) with relative phases of $\pi$ in the case of  $\widetilde{V}_+$ and $0$ for $\widetilde{V}_-$. This leads to
\begin{subequations}\label{lipar}
\begin{align}
&\widetilde{V}_+(\xi, \phi,\zeta)=i\,\frac{\xi\sin\phi}{(1+i\zeta)^2}\, e^{-\frac{\xi^2}{1+i\zeta}},\label{linparp}\\
&\widetilde{V}_-(\xi, \phi,\zeta)=\frac{\xi\cos\phi}{(1+i\zeta)^2}\, e^{-\frac{\xi^2}{1+i\zeta}}. \label{linparm}
\end{align}
\end{subequations} 
Both these functions satisfy the scalar paraxial equation. With this form one gets the following  components of the electric field  
\begin{subequations}\label{parline}
\begin{align}
\widetilde{\mathcal{E}}_{\xi_x}=&-4\varepsilon^2\xi_x\xi_y\partial_\zeta\left[\frac{1}{(1+i\zeta)^3}\, e^{-\frac{\xi^2}{1+i\zeta}}\right],\label{parlinex}\\
\widetilde{\mathcal{E}}_{\xi_y}=&\;2 i \left(1-\frac{\xi^2}{1+i\zeta}\right)\frac{1}{(1+i\zeta)^2}\,\, e^{-\frac{\xi^2}{1+i\zeta}}\nonumber\\
&+2\varepsilon^2(\xi_x^2-\xi_y^2)\partial_\zeta\left[\frac{1}{(1+i\zeta)^3}\, e^{-\frac{\xi^2}{1+i\zeta}}\right],\label{parliney}\\
\widetilde{\mathcal{E}}_{\zeta}=&\;4i\varepsilon \xi_y\partial_\zeta\left[\frac{1}{(1+i\zeta)^2}\, e^{-\frac{\xi^2}{1+i\zeta}}\right].\label{parlinez}
\end{align}
\end{subequations}
The $\xi_x$ and $\zeta$ components turn out to be small due to the coefficients $\varepsilon$ and $\varepsilon^2$, so the field is aligned with the $\xi_y$ axis, with tiny deviations which are visible on the first diagram of Fig. \ref{Elin}. This conclusion is true only within the circle of rough radius equal to $1$ due to the factor $\left(1-\frac{\xi^2}{1+i\zeta}\right)$. Outwards, the field orientation gets reversed.

The alignment with the $\xi_x$ axis can be achieved in an analogous way with the choice
\begin{subequations}\label{lipara}
\begin{align}
&\widetilde{V}_+(\xi, \phi,\zeta)=i\,\frac{\xi\cos\phi}{(1+i\zeta)^2}\, e^{-\frac{\xi^2}{1+i\zeta}},\label{linparap}\\
&\widetilde{V}_-(\xi, \phi,\zeta)=-\frac{\xi\sin\phi}{(1+i\zeta)^2}\, e^{-\frac{\xi^2}{1+i\zeta}}. \label{linparam}
\end{align}
\end{subequations} 
The result is is visualized on the right diagram of Fig. \ref{Elin}.

\subsection{$\varepsilon^2$ terms with nonzero OAM}\label{wpp}

When inspecting formulas (\ref{parpee}) and (\ref{parpbe}) -- naturally with $\beta=1$ -- one observes that the $\varepsilon^2$ terms are accompanied by the derivative with respect to $\zeta$. If the potentials $\widetilde{V}_\pm$ represent the beams with OAM, these term will get the additional factor of order $(n+1)$ ($n$ being the vortex index). This is clearly evident in the expressions for the potentials, which in such a case can have, for example, a Gaussian form:
\begin{equation}\label{vpp}
\widetilde{V}_\pm(\xi, \phi,\zeta)\sim\frac{\xi^ne^{i n\phi}}{(1+i\zeta)^{n+1}}\, e^{-\frac{\xi^2}{1+i\zeta}}.
\end{equation}
This is because the Gouy phase varies proportionally to $(n+1)$ along the beam. This implies that in beams with non-zero OAM the additional terms referred to are expected to play a more significant role. On the other hand, by setting the value of $n$ too large the right hand sides of Maxwell's equations (\ref{romax}) are also magnified, which in turn renders the paraxial approximation less justified. For this reason, we will examine below the effect of epsilon expressions on the polarization for $n=1,2,3$.

For concreteness let $\widetilde{V}_+$ be chosen in the form of (\ref{vpp}), and $\widetilde{V}_-=-\widetilde{V}_+$. With this choice the effect is relatively well visible. For simplicity let us first look at these expressions  in the focal plane, i.e., for $\zeta=0$. One then gets the transverse cylindrical components of the electric field:
\begin{subequations}\label{upe}
\begin{align}
&\widetilde{\mathcal{E}}_\xi=\xi^{n-1}e^{-\xi^2}\Big[n(1-\xi)-2\xi^2\label{upex}\\
&\;-\varepsilon^2\left(n(n+1)(1+\xi)-\xi^2(4+3n+n\xi-2\xi^2)\right)\Big]\cos n\phi,\nonumber\\
&\widetilde{\mathcal{E}}_\phi=\xi^{n-1}e^{-\xi^2}\Big[n(1-\xi)-2\xi^2\label{upef}\\
&\;+\varepsilon^2\left(n(n+1)(1+\xi)-\xi^2(4+3n+n\xi-2\xi^2)\right)\Big]\sin n\phi.\nonumber
\end{align}
\end{subequations}
Attention should be paid to factors $n(n+1)\varepsilon^2$, that quickly increase with $n$. For instance for $n=2$ at a particular exemplary point of $\xi=0.7$, $\phi=\pi/5$, one has
\begin{equation}
\widetilde{\mathcal{E}}_\xi\approx -0.05\cdot(1+13.4\,\varepsilon^2),\qquad \widetilde{\mathcal{E}}_\phi\approx-0.15\cdot(1-13.4\,\varepsilon^2),\label{upen}
\end{equation}
which proves that the role of the second term may not be marginal. 

\begin{figure}[h!]
\begin{center}
\includegraphics[width=0.48\textwidth,angle=0]{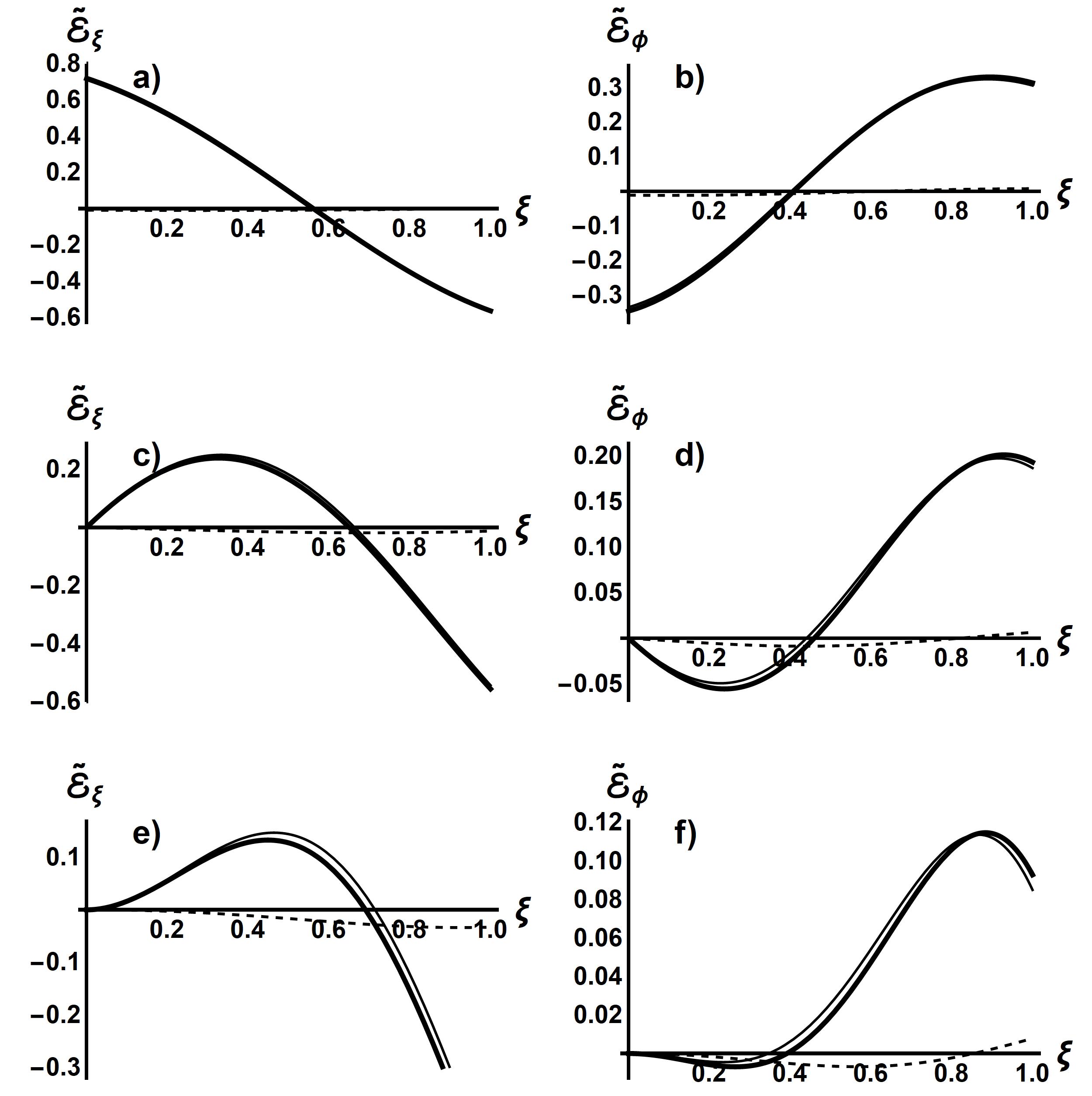}
\end{center}
\vspace{-4.5ex}
\caption{The radial [plots a), c) and e)] and azimuthal [plots b), d) and f)] components of the electric field for $\varepsilon=0.1$, and $\phi=\pi/5$ in the plane $\zeta=0.5$ with increasing value on $n$ ($n=1,2,3$ from top to bottom). The dashed line represents the $\varepsilon^2$ correction, the thin line is the field without this correction and the thick line is the sum of both.}
\label{fields1}
\end{figure}

\begin{figure}[h!]
\begin{center}
\includegraphics[width=0.48\textwidth,angle=0]{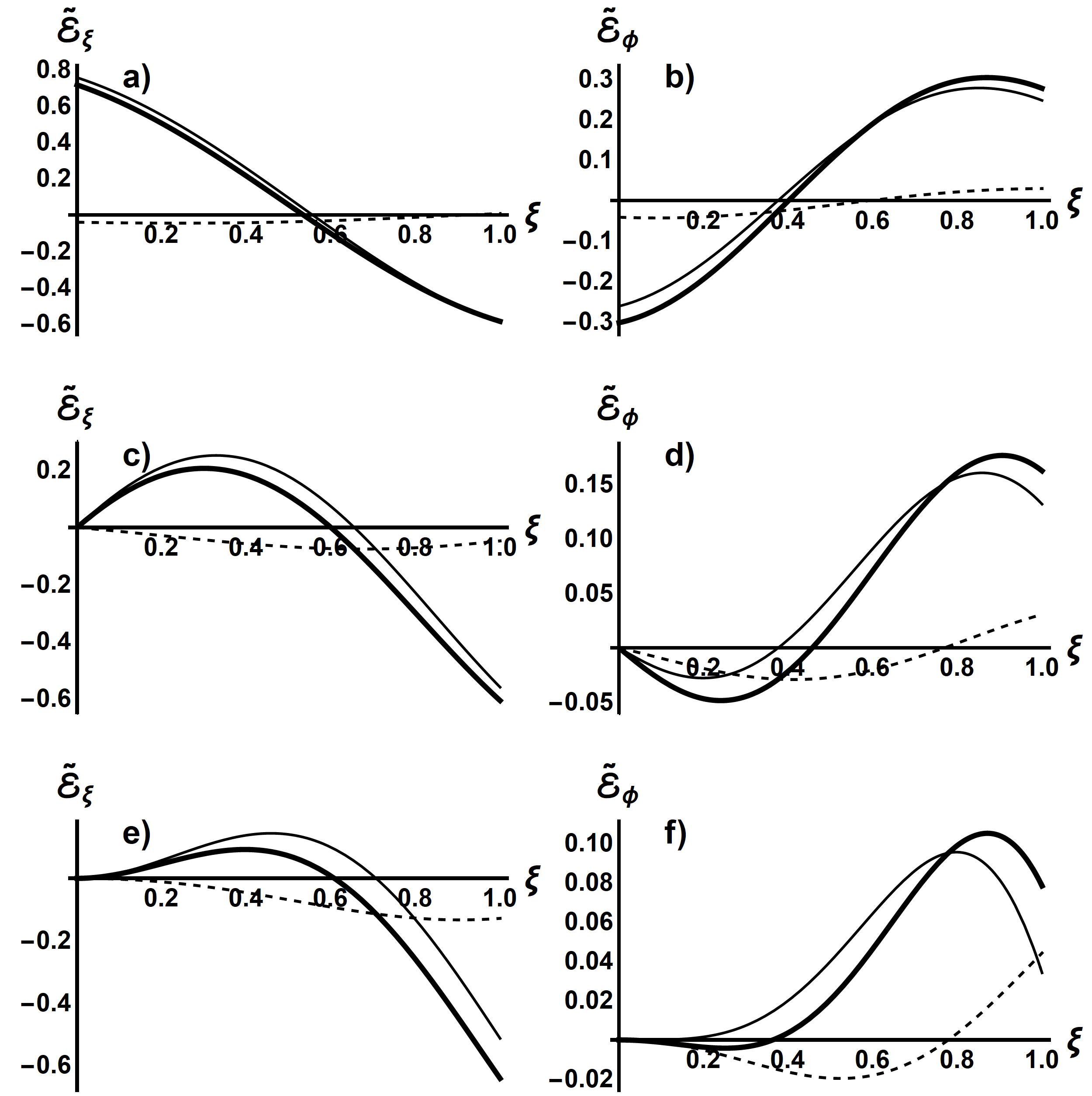}
\end{center}
\vspace{-4.5ex}
\caption{Same as Fig. \ref{fields1} but for $\varepsilon=0.2$.}
\label{fields2}
\end{figure}

Figures \ref{fields1} and \ref{fields2} demonstrate the deviations of the radial and azimuthal components of the electric field from those without $\varepsilon^2$ terms in the plane $\zeta=0.5$ and $\phi$ set to $\pi/5$. For $\varepsilon=0.1$ this deviation is tiny for $n=1$, but becomes  essential starting from $n=3$. If the value of $\varepsilon$ reaches $0.2$, strong deviations are observed already for $n=1$.

\begin{figure}[h!]
\begin{center}
\includegraphics[width=0.48\textwidth,angle=0]{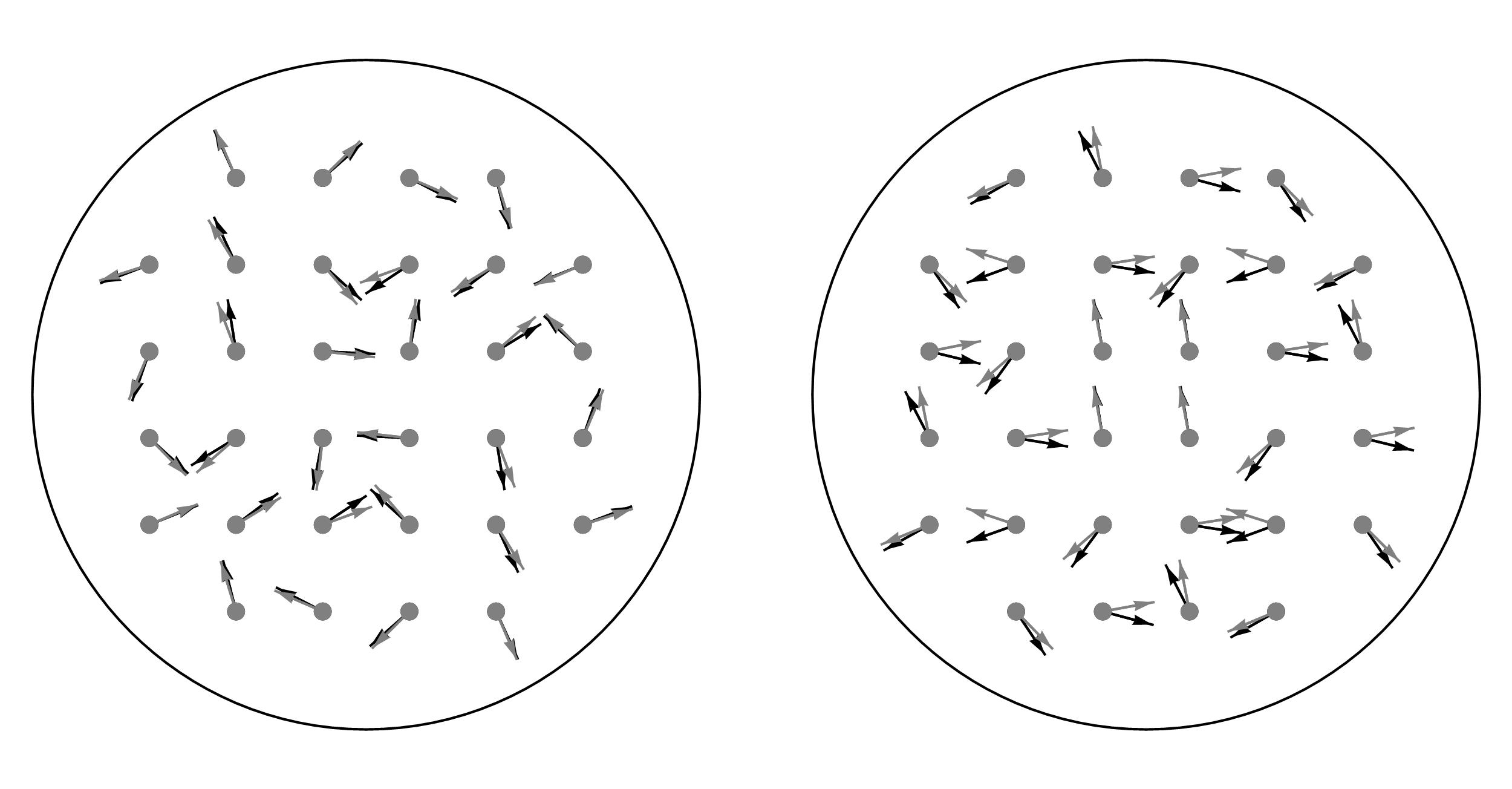}
\end{center}
\vspace{-4.5ex}
\caption{The local transverse electric field without the corrections (grey arrows), and corrected with $\varepsilon^2$ terms (black arrows) for $n=1$ (left diagram) and $n=2$ (right diagram). The values of other parameters are: $\varepsilon=0.2$, $\zeta=0.5$, and the circle radius corresponds to $1.1w_0$.}
\label{Earrp}
\end{figure}

\begin{figure}[h!]
\begin{center}
\includegraphics[width=0.48\textwidth,angle=0]{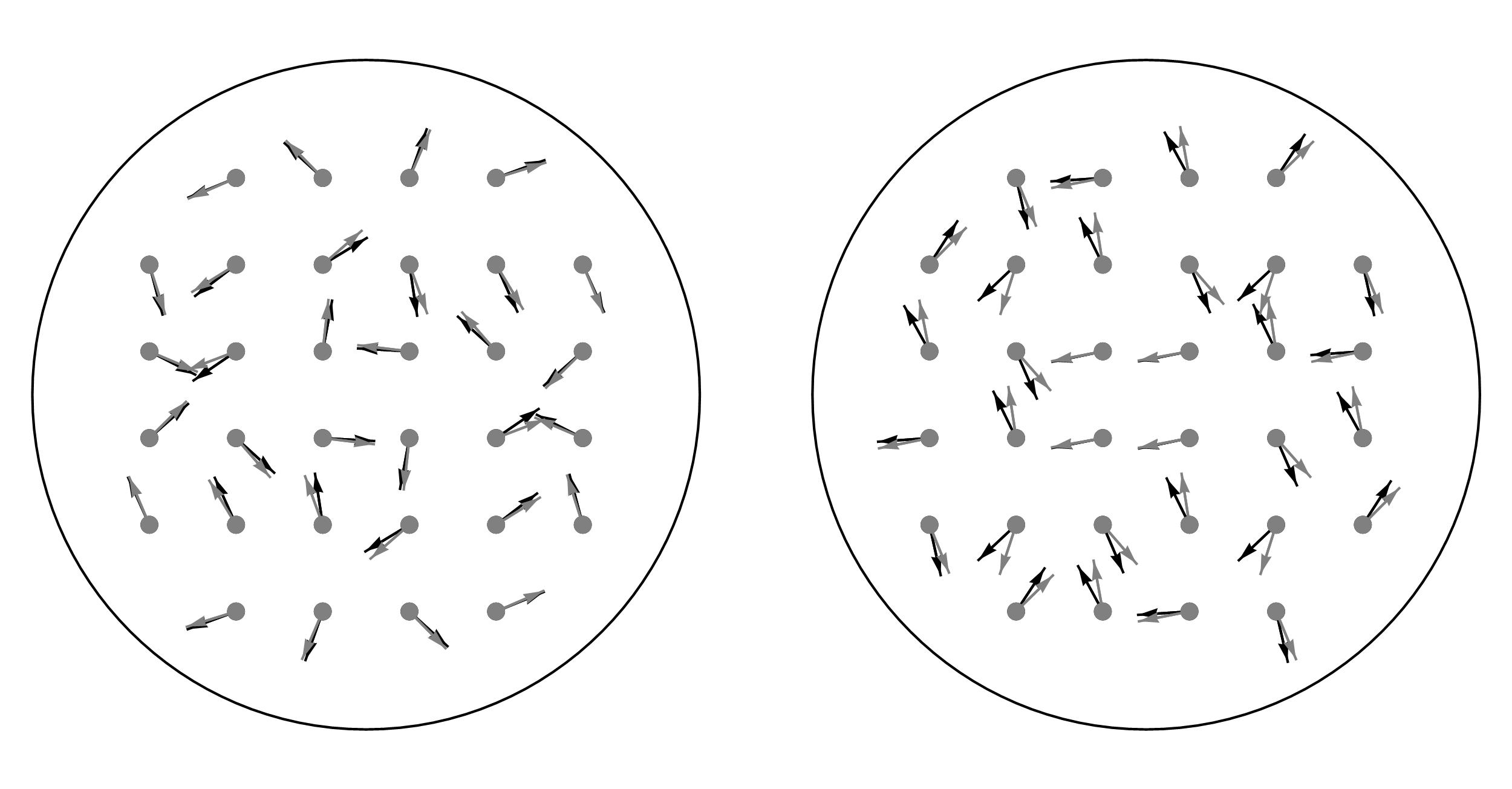}
\end{center}
\vspace{-4.5ex}
\caption{Same as Fig. \ref{Earrp} but for magnetic field.}
\label{Barrp}
\end{figure}

At least in the focal plane the $\varepsilon^2$ contributions to the radial and azimuthal components merely differ in sign (i.e., are similar in magnitude), so that at larger values of $\varepsilon$ (e.g. $0.1-0.2$) both can change substantially due to the $\varepsilon^2$ corrections. Some examples of this effect are shown in Figs~\ref{Earrp} and~\ref{Barrp} for electric an magnetic fields respectively.

\section{Concluding remarks}\label{sum}

A precise description of vector paraxial beams can be of fundamental importance for modern applications in high-resolution microscopy, particle trapping or optical lithography. Therefore, the motivation of the present work was to establish the form of the paraxial Maxwell equations and then to find their {\em exact} solutions in the form of explicit expressions for the components of electromagnetic fields. It has been shown that the derived fields satisfy full, non-approximated Maxwell equations up to the order of $(\lambda/w_0)^2$. Although it is also essential to find beyond-paraxial corrections, which is often done, it is primarily the paraxial beam itself that should be found with best accuracy.

The paraxial solutions provided in~\cite{patta,es,hil95,que,nomoto,pastor}, satisfy the full Maxwell equations up to the lower order, i.e., that of $\lambda/w_0$. The exception is the work \cite{lew}, where the authors claim that the formulas obtained describe the improved paraxial solutions, however, the expressions are quite intricate and the relevant estimations are not explicitly presented in the work.

The implications of the additional expressions for the local electric field have been demonstrated graphically for Gaussian beams with non-zero orbital angular momentum, as well as for a superposition of beams that exhibits quasi-linear polarization. Although they are not too large, they can be crucial for the local polarization and thus for the precision required in modern high-resolution optics whose applications necessitate the use of tight beams. It is enough to mention that the size of the components in modern processors is expressed in nanometers, and such precision is also required, for example, when observing biological structures at the molecular level. An appropriate choice of potentials $\widetilde{V}_\pm$ enables the study of electric and magnetic fields for wide class of paraxial beams.

\end{document}